\documentclass[preprint]{elsarticle}

\journal{Journal of Theoretical Biology}

\usepackage[utf8]{inputenc} 
\usepackage[T1]{fontenc}    
\usepackage{hyperref}       
\usepackage{url}            
\usepackage{booktabs}       
\usepackage{amsfonts}       
\usepackage{amsmath}       
\DeclareMathOperator*{\argmax}{arg\,max} 
\DeclareMathOperator*{\argmin}{arg\,min} 
\usepackage{nicefrac}       
\usepackage{microtype}      
\usepackage{algorithm}
\usepackage{algorithmic}
\usepackage{graphicx}
\usepackage{fancyvrb}
\usepackage[margin=1in]{geometry}
\usepackage{xcolor}
\usepackage{lineno}
\usepackage{fixltx2e}
\definecolor{mygray}{RGB}{120,120,120}
\definecolor{lightblue}{RGB}{0,176,240}
\usepackage[most]{tcolorbox}

\newcommand{\hl}{}

\begin{document}

\begin{frontmatter}

\title{\textbf{Understanding Memory B Cell Selection}}

\author[1]{Stephen Lindsly}
\author[2]{Maya Gupta\corref{corrauth}}
\ead{relativeentropy@gmail.com}
\author[1]{Cooper Stansbury}
\author[1,3]{Indika Rajapakse\corref{corrauth}}
\ead{indikar@umich.edu}
\cortext[corrauth]{Corresponding Authors}
\address[1]{Department of Computational Medicine and Bioinformatics, University of Michigan, Ann Arbor}
\address[2]{Google Research, Mountain View, CA}
\address[3]{Department of Mathematics, University of Michigan, Ann Arbor}

\begin{abstract}
\noindent The mammalian adaptive immune system has evolved over millions of years to become an incredibly effective defense against foreign antigens. The adaptive immune system's humoral response creates plasma B cells and memory B cells, \hl{each with their own immunological objectives.} The affinity maturation process is widely viewed as a heuristic to solve the global optimization problem of finding B cells with high affinity to the antigen. However, memory B cells appear to be purposely selected earlier in the affinity maturation process and have lower affinity. We propose that this memory B cell selection process may be an approximate solution to two optimization problems: optimizing for affinity to similar antigens in the future despite mutations or other minor differences, and \hl{optimizing to warm start the generation of plasma B cells in the future.} We use simulations to provide evidence for our hypotheses, taking into account data showing that certain B cell mutations are more likely than others. \hl{Our findings are consistent with memory B cells having high-affinity to mutated antigens, but do not provide strong evidence that memory B cells will be more useful than selected naive B cells for seeding the secondary germinal centers.}
\end{abstract}



\end{frontmatter}


\section{Introduction}
The immune system is an effective \hl{threat mitigation} system that deploys a number of \hl{learned identification algorithms}. While the \emph{innate immune system} is adept at identifying foreign invaders, or antigens, it must engage the \emph{adaptive immune system} to create a more massive and specific response.  A core aspect of the adaptive immune system's humoral response is training two types of \hl{B cells} through a process called affinity maturation (AM): plasma B cells which generate antibodies to identify the current antigen, and memory B cells which are used in subsequent immune responses to \hl{identify} similar antigens in the future. \hl{The AM process is highly unusual, in that a specific region of DNA within participating B cells is mutated to generate offspring which are selected to have higher affinity to the antigen in question.} The preservation of DNA sequences is usually of utmost importance in most cells, but the region of the genome which defines the shape of the B cell receptor must be \hl{rapidly} modified for the B cell receptor to have a chance of becoming better at recognizing the antigen of interest \cite{meyer2012theory}. These mutations are responsible for the B cells' incredible ability to recognize practically any antigen that they are presented, making the mammalian adaptive immune system one of the most effective
\hl{learned identification systems} in the natural world.                             

\hl{In this paper, we consider whether the plasma B cell and memory B cell generation processes can be interpreted as trying to satisfy specific objectives, and if so, can we state these objectives precisely?} We borrow standard \hl{ideas} from machine learning, where it is common to first specify 
\hl{an ideal} mathematical objective to be optimized (such as minimizing the expected error rate of \hl{a learned identification system),} then propose heuristic algorithms that approximately optimize that mathematical objective. Similarly, we hypothesize that due to evolutionary pressures, the AM processes \hl{act like heuristics that approximately optimize for idealized immunological objectives.} We hypothesize what those \hl{evolutionarily-adaptive immunological objectives} might be, then compare how well different B cells \hl{satisfy these objectives} when faced with adversarially-mutated antigens via simulations. These findings lead us to propose new hypotheses about the implicit \hl{objectives} of the immune system's training of naive B cells to become memory B cells. 

First in Section \ref{sec:plasma}, we \hl{review how naive B cells are recruited and trained to become plasma B cells, and present a hypothesis for the objectives of this training mathematically. In Section }\ref{sec:memory}\hl{, we consider the mathematical optimization objectives of the more enigmatic training process that leads to the generation of memory B cells. We define \textit{plasma B cell training} or \textit{training plasma B cells} as the process of generating plasma B cells from naive B cells during AM. Similarly, we define \textit{memory B cell training} or \textit{training memory B cells} as the process of generating memory B cells from naive B cells during AM.} We test our hypotheses via simulations in Section \ref{sec:simulations}, and conclude with a discussion of open questions in Section \ref{sec:conclusions}.

\begin{figure}[t]
    \centering
    \includegraphics[width=.9\textwidth]{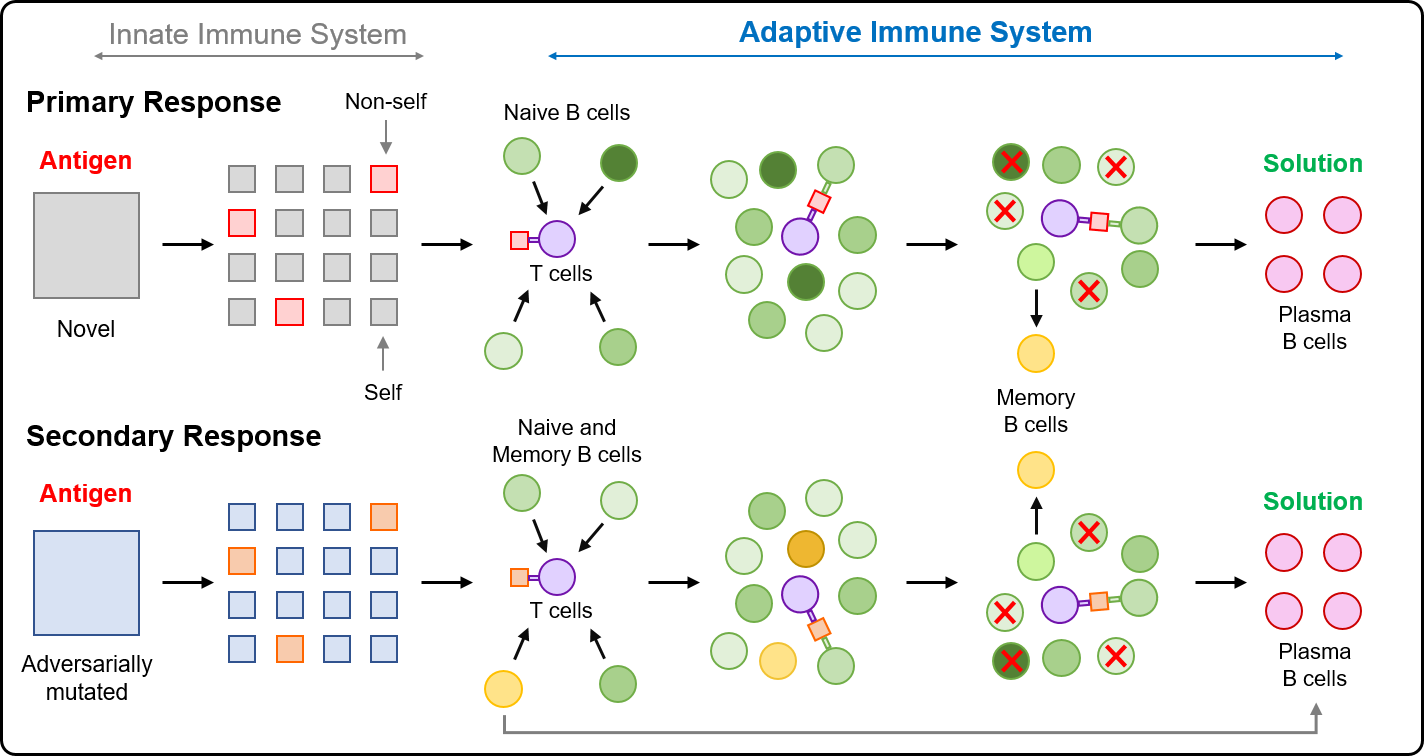}
    \caption{High-level illustration of the adaptive immune system. First, an antigen enters the body, then the innate immune system identifies pieces of the antigen as non-self.  (Top:) The adaptive immune system’s responds to the identified antigen. A diverse set of random naive B cells that have some initial affinity to the antigen flock together and form a germinal center \cite{meyer2012theory,tas2016visualizing}. These B cells proliferate and mutate when selected by T\textsubscript{FH} cells for their affinity to the antigen. B cells with moderate affinity are stored for later use as memory B cells. High affinity B cells differentiate into plasma B cells, which are the solution to a particular antigen. (Bottom:) A mutated version of a previously encountered antigen, or an antigen from a related pathogen, is presented to the adaptive immune system. It responds by forming germinal centers with both random naive B cells (with some initial affinity to the  antigen) and memory B cells from the first encounter. \hl{Memory B cells can also be used directly by differentiating into plasma B cells}.}
    \label{fig:cartoon}
\end{figure}

\section{Plasma B cell training} 
\label{sec:plasma}
We review the AM process that trains naive B cells to become plasma B cells, then consider what mathematical criteria the plasma B cell training may have evolved to optimize. 

\subsection{Affinity Maturation Of Plasma B Cells}
AM begins by recruiting naive B cells with some initial affinity to the antigen to secondary lymphoid organs. These naive B cells, along with T follicular helper (T\textsubscript{FH}) cells and follicular dendritic cells, concentrate into temporary structures known as germinal centers (Fig. \ref{fig:cartoon}A) \cite{meyer2012theory,tas2016visualizing}. Germinal centers (GCs) ensure that these cells are in close proximity, \hl{thus facilitating rapid mutation and evaluation of B cells receptor sequences. During AM, B cells are evaluated by T}\textsubscript{FH} \hl{cells for their affinity to the antigen through the length of interaction between them, based on  antigen presentation by B cells} \cite{mesin2016germinal,murphy2016janeway}. If the initial affinity of a B cell is high, it receives a chemical signal from the T\textsubscript{FH} cell to move to a separate area of the GC and proliferate. \hl{While the B cell is proliferating, a specific section of the genome called the hypervariable region is exposed to an enzyme, activation-induced cytidine deaminase (AID)} \cite{muramatsu2000class,bannard2017germinal}. AID is able to deaminate cytosine creating uracil, a nucleotide that is not normally found in DNA. The operation that repairs these changes is error-prone, leading to mutations in the DNA sequence \cite{martin2015somatic}.

The process of deamination and mutations during repair is referred to as \emph{somatic hypermutation} (SHM). After proliferating, the B cells return to the area of the GC containing T\textsubscript{FH} cells and are reevaluated for their affinity towards the antigen. This iterative process of proliferation, mutation, and affinity evaluation continues until the B cells have a sufficiently high affinity to the antigen. At this point, the B cells differentiate into plasma B cells and begin to produce antibodies which allow for the immune system to eradicate the antigen.

\subsection{Affinity Maturation As An Algorithm}
We model AM in Algorithm \ref{alg:shm}, which we use to simulate AM in our experiments.  We simplify a few known or uncertain issues about AM, detailed in Subsection \ref{sec:knownIssues}. 

Each \emph{naive} B cell \hl{receptor sequence} is generated randomly by a combinatorial mix of its V, D, J, and C gene segments, as well as through junctional diversity between these segments \cite{sompayrac2019immune}. Naive B cells span at least 100 million possibilities \cite{sompayrac2019immune}. The naive B cells recruited to a germinal center are \hl{cells} that already have some promising affinity $s$ to the antigen $a$. \hl{SHM then mutates nucleotides in a region of the DNA sequence approximately 100 base pairs (bp) long which contributes to the definition of the B cell receptor structure }\cite{martin2015somatic}. Mutations in the hypervariable region are on the scale of $10^6$ times more likely than mutations outside of this region \cite{martin2015somatic}.  Mutations can be swaps, insertions and deletions in a categorical space modeled as $\{A, T, C, G, \emptyset \}^{100}$, where $\emptyset$ connotes a deletion. \hl{It should be noted that although the combinatorial space of possible mutations is large, many specific mutations immediately lead to apoptosis.} AM optimization is parallelized and distributed over $G$ germinal centers, which algorithmically can be thought of as $G$ different parallel processors. We model the different germinal centers as working independently, though there may be biochemical signaling between them. In practice, \hl{an organism trains for multiple independent antigens simultaneously, but for simplicity, we consider one antigen at a time.}

\begin{figure}[H]
\begin{algorithm}[H]
\caption{Affinity Maturation Algorithm for Training Plasma and Memory B Cells}
\label{alg:shm}
\begin{algorithmic}[1]
\REQUIRE{an antigen $a \in \mathcal{A}$}
\REQUIRE{an affinity score $s(b, a) \rightarrow \mathbf{R}$ for B cell $b \in \mathcal{B}$ and antigen $a \in \mathcal{A}$ } 
\REQUIRE{a low affinity threshold $\epsilon$ to enter a germinal center}
\REQUIRE{a high affinity threshold $\tau >> \epsilon$ to become a plasma B cell}
\REQUIRE{a die-off rate $d \in [0,1]$ for germinal center B cells}
\REQUIRE{probability $p$ that a B cell is measured by a T\textsubscript{FH} cell}
\REQUIRE{probability function $q(t, s)$ of producing a memory B cell on iteration $t$ given affinity score $s$ that is monotonically increasing in $s$, and might be monotonically decreasing or unimodal in $t$} 
\REQUIRE{probability function $r(s)$ of a proliferation signal from a T\textsubscript{FH} cell after affinity measurement}
\REQUIRE{a Bernoulli random number generator Bernoulli($p$) that outputs 1 with probability $p$ and 0 otherwise}
\STATE \textbf{initialize} the set of plasma B cells $B^* = \emptyset$ and the set of memory B cells $V^* = \emptyset$
\FOR{$g = 1, \ldots, G$ \textrm{ germinal centers} }
\STATE sample an initial set of $J_g$ naive B cells $B_g^0$ such that $s(b, a) > \epsilon$ for all $b \in B_g^0$ 
\FOR{$t = 1, \ldots, T$ \textrm{ iterations} }
\FOR{$b \in B_g^t$}

\IF{Bernoulli($p$) == 1}
\STATE the B cell $b$ is observed by some nearby T\textsubscript{FH} cell which measures $s(b, a)$
\IF{$s(b, a) \geq \tau$}
\STATE insert $b$ into the set of plasma B cells $B^*$ such that $b \in B^*$ 
\STATE \textbf{break}
\ENDIF

\IF{Bernoulli($q(t, s(b, a))$) == 1}
\STATE insert $b$ into the set of memory B cells $V^*$ such that $b \in V^*$
\STATE \textbf{break}
\ENDIF

\IF{Bernoulli($r(s(b, a))$) == 1}
\STATE proliferate: B cell $b$ sent to divide and mutate some number of times, and its mutated copies are added to the set $B^{t+1}_g$
\STATE \textbf{break}
\ENDIF

\IF{Bernoulli($d$)}
\STATE $b$ dies
\STATE \textbf{break}
\ENDIF

\ENDIF 

\STATE $b$ is added to the set $B^{t+1}_g$

\ENDFOR 
\ENDFOR 
\ENDFOR 
\STATE \textbf{return}{ the set of plasma cells $B^*$ and the set of memory B cells $V^*$ }
\end{algorithmic}
\end{algorithm}
\end{figure}

\subsection{Known Simplifications Of Algorithm \ref{alg:shm}}\label{sec:knownIssues}
We note that Algorithm \ref{alg:shm} simplifies a few known \hl{characteristics of AM. We believe these simplifications are minor and that they do not affect the major conclusions of this work.}

\hl{We model the algorithm as $T$ discrete iterations but in practice, AM is continuous process that is partly time-limited because of antigen decay and external pressures. However, to our knowledge, there is not an exact limit to the number of divisions that can occur or an exact timeline that must be met during AM. Furthermore, there is a chance that the immune system does not find a solution fast enough, causing the host to die. While we recognize that this occurs in the natural system, it is not the focus of our simulation. Therefore, we use a fixed number of iterations as an approximation of the time limits the real immune system faces. In addition, we identify the highest affinity B cells at the end of our simulation as plasma B cells for simplicity, but plasma B cells are not selected simultaneously at the end of AM and may not have the absolute highest affinity.}

Before SHM begins, the initial \hl{B cell population may have undergone undirected proliferation,} which means the initial random sample may be better modeled as random clusters of B cells. Algorithm \ref{alg:shm} allows the germinal centers to grow without bounds, though the die-off rate $d$ will tend to \hl{limit the population size in the germinal centers.} In practice, the size of germinal centers are also bounded by physical volume constraints and biochemical resource constraints. \hl{We use a constant die-off rate $d$, but there is some evidence die-off probability decreases as affinity increases }\cite{Anderson:09}. \hl{We include simplified version of this behavior by biasing the death probabilities of B cells in the germinal center towards lower affinity cells.}

T\textsubscript{FH} cells measure \emph{nearby} B cells for their affinity, so there is only some probability that a specific B cell will have its affinity measured, and that probability a B cell's affinity gets measured is thought to be a function of spatial organization (which is indirectly affected by affinity) and direct affinity. The affinity \hl{and spatial proximity} of a particular B cell influences its likelihood to \hl{be selected and induced to proliferate. We do not consider the spatial organization between T}\textsubscript{FH} \hl{cells and B cells explicitly. The strength of the proliferation signal is generally proportional to the affinity of the B cell, such that a high affinity B cell is more likely to proliferate many times before returning for another iteration of affinity evaluation }\cite{mesin2016germinal,de2015dynamics}. \hl{Similar to the increased likelihood of death for a low affinity B cell, we bias the selection of B cells for proliferation towards higher affinity B cells.}

Algorithm \ref{alg:shm} may oversimplify AM in other ways as well that we are not aware of, or that are not yet known. 

\subsection{Plasma B Cells Are Created To Optimize Antigen Affinity Given Limited Time}
AM is a process that has long been framed as the immune system acting as a global optimization algorithm trying to find a B cell that best identifies a given antigen through SHM \cite{Theodosopoulos:2002}. That is, AM acts as if it were a heuristic to solve,
\begin{equation}
    \argmax_{b \in \mathcal{B}} s(b, a)  
    \label{eqn:plasmaOpt}
\end{equation}
where $a$ is a given antigen, $b$ is a B cell from the set $\mathcal{B}$ of all possible B cells, and $s$ is the affinity function that models the quality of the lock-and-key physical and biochemical interaction of $b$ and $a$. Note that $\mathcal{B}$ is a very large categorical space defined by the variable-length DNA sequence that encodes the B cell receptor.  

However, the objective (\ref{eqn:plasmaOpt}) does not recognize the fact that a plasma B cell does not need to be a perfect match to the antigen. In fact, there appears to be a sufficient affinity $\tau$ such that once an affinity of $\tau$ is reached, the B cell \hl{is induced to differentiate into a plasma B cell.} Further, the immune system is under time pressure to produce such sufficiently high-affinity plasma B cells as fast as possible.

Therefore, we propose that a more \hl{realistic} model of what the plasma B cell generation process is optimizing should also depend on the sufficient affinity $\tau > 0$, and the given set $\mathcal{B}_0$ of initial naive B cells in the germinal center.  To capture the time pressure, \hl{we model probabilistic mutations to B cells in the germinal center at each discrete time iteration.} Given a naive B cell $b \in \mathcal{B}_0$, let $M(b) \in \mathcal{B}$ be a new \hl{random B cell produced by a single random mutation of $b$.} Let $M^K(b) = M(M(\ldots(M(b))\ldots))$ denote the random B cell generated after $K$ random mutations of $b$, so that the random B cell $M^K(b)$ can be any of the B cells reachable by $K$ mutations $M$ of the initial B cell $b$, \hl{with the corresponding probabilities dependent on the sum of the likelihood of the different mutations paths that could produce $M^K(b)$ starting from $b$.}  

Then we hypothesize the plasma B cell selection process is a heuristic evolved to minimize the number of of mutations $K \in \mathcal{N}$ needed so that on average $K$ random mutations will produce at least one B cell in the germinal center with sufficient affinity $\tau$ to the antigen:
\begin{align}\label{eqn:minExpectedNumMutationsPlasmaCells}
    \min K \textrm{ subject to } \Bigg( E_{M^K} \bigg[ \max_{b \in \mathcal{B}_0} s(M^K(b), a) \bigg]  \Bigg) \geq \tau, 
\end{align}
where $E[\cdot]$ is the standard expectation operator (average) with respect to the random variable's possible outcomes weighted by their probabilities. This objective is consistent with our Algorithm \ref{alg:shm}. 

Clearly, the immune system is not a sentient entity that \hl{explicitly formulates the criteria } (\ref{eqn:minExpectedNumMutationsPlasmaCells}) \hl{and subsequently identifies a heuristic to optimize it.} Rather, our hypothesis is that evolutionary pressures have selected for a plasma B cell generation process that better optimizes (\ref{eqn:minExpectedNumMutationsPlasmaCells}).

\section{Memory B Cell Training} \label{sec:memory}
Similar to plasma B cells, memory B cells are created within the germinal center, but there are key differences in the generation of these two cell types to achieve their respective objectives \cite{mesin2016germinal,Weisel:2017,suan2017plasma}. We first review how memory B cells are created, and then consider what criteria they are optimized for, analogous to our criteria (\ref{eqn:minExpectedNumMutationsPlasmaCells}) for plasma B cells.

\subsection{Background On Memory B Cells}
While the name \emph{memory B cell} may invoke the idea that a memory B cell is long-term copy of a plasma B cell, the truth is more complicated. Memory B cells do not undergo the entire AM process like plasma B cells do. In fact, memory B cells are characterized by their relatively low affinity compared to plasma B cells, and low SHM load (number of mutations gathered from SHM) \cite{suan2017plasma}. This implies that while memory B cells have initially high affinity to an antigen relative to the naive B cell repertoire, they do not undergo AM to the extent of plasma B cells, \hl{and tend to have lower affinity to the current antigen than plasma B cells.}

The gene \textit{BACH2} plays an important role in the development of memory B cells within the germinal center, and in their eventual differentiation \cite{suan2017plasma, shinnakasu2016regulated}. \textit{BACH2} has been found to be inversely correlated with the help a B cell receives from T\textsubscript{FH} cells, and  the resulting weak interactions with T\textsubscript{FH} cells allow for \textit{BACH2} expression to remain high. Critically, the relationship between \textit{BACH2} expression and T\textsubscript{FH} cells allows for \emph{some} help from T\textsubscript{FH} cells in order for cell survival within the germinal center, but prevents the B cell precursor from entering the \hl{area} where it would proliferate and mutate via SHM. This leads to three subsets of B cells within the germinal center: (1) high affinity B cells which are selected for by T\textsubscript{FH} cells to proliferate and mutate via SHM, and eventually lead to plasma B cell differentiation, (2) moderate affinity B cells (low compared to plasma B cell precursors, high compared to the average naive B cell) whose selection by T\textsubscript{FH} cells is tempered by \textit{BACH2}, leading to memory B cells, and (3) low affinity B cells which receive little or no help from T\textsubscript{FH} cells leading to apoptosis \cite{mesin2020restricted,taylor2015apoptosis}. 

\hl{Memory B cells are similar to naive B cells in terms of their transcriptional profiles, which enables them to circulate freely within the organism and to monitor for future instances of antigens.} Despite these similarities, they exhibit over-expression of anti-apoptotic genes which allows for the memory B cell to live for extraordinarily long periods of time and therefore the ability to recognize antigens in the future \cite{suan2017plasma}.  

\subsection{What Are Memory B Cells Optimized To Do?}\label{sec:memoryTraining}
If the immune system's \hl{objective were rote memorization of the highest affinity B cell receptors to the current antigen $a$, we might expect the memory B cell receptor repertoire to be nearly identical to the plasma B cells receptor repertoire, but they are not.} One might alternatively expect AM to take advantage of the luxury of time it has before it needs the memory B cells \hl{to mutate more so that the memory B cells could have even higher affinity $s(b,a)$ to the antigen than the plasma B cells, further optimizing }(\ref{eqn:plasmaOpt}). That also does not appear to \hl{be the case.} While both those options should be biologically feasible, the immune system does something radically different to create memory B cells: it selects memory B cells \emph{earlier} in AM than plasma B cells, and thus the memory B cells on average have \emph{lower} affinity $s(b,a)$ than plasma B cells. 

\hl{To explain why the memory B cells are so poorly-fit to the current antigen $a$,  we propose two hypotheses for the objective function that memory B cells may be heuristically trying to optimize.} Our two hypotheses follow from the dual role of memory B cells \cite{Dogan:2009,Weisel:2017}. First, when \emph{future} incarnations of the antigen $a$ attack, the memory B cells are used as-is to differentiate into plasma B cells and eradicate the \emph{mutated} antigen. Second, memory B cells are used to warm start AM's secondary training of plasma B cells.  In fact, recent evidence shows that a large portion of the plasma B cells in the secondary response are memory B cells from the first response, and some memory B cells are also used to seed the new germinal centers to optimize secondary response plasma B cells \cite{mesin2020restricted}.

\subsection{Training For Affinity To A Mutated Antigen}\label{sec:diversitygood}
We propose that the key issue for memory B cells is that  \hl{the future instance of the antigen they must mitigate is almost certainly a mutation $\tilde{a}$ of the original antigen $a$. At the time the memory B cells are created, the future mutated antigen $\tilde{a}$ is unknown, but we can characterize it as a randomly mutated antigen $\tilde{A}$. In this paper, we use the standard probability notation that a capital letter denotes a random variable, and its corresponding lower-case letter denotes the realization of that random variable. For example, if you roll a six-sided die, the random value $X \in \{1,2, \ldots, 6\}$ refers to the die roll before you look at it because at that point you only know the probability of its six values, but once you see the die roll, it is a deterministic value $x \in \{1,2,\ldots, 6\}$.} 

Let $\tilde{A}$ be a random antigen drawn from some conditional probability distribution \hl{$P_{\tilde{A}|a}$ that depends on the current antigen $a$, and models the probability of possible future mutations to $a$ and the probability that such a mutation is presented to the host organism within the lifespan of memory B cells selected during the immune response to  $a$. If only a single memory B cell were required for a secondary response, a logical generalization of }(\ref{eqn:plasmaOpt}) \hl{would be to select a memory B cell that will, on average, have high affinity to the random mutated antigen $\tilde{A}$. }

\begin{equation}
\argmax_{b \in \mathcal{B}} \: \: E_{\tilde{A}} \left[ s(b, \tilde{A}) \right].
\label{eqn:opt}
\end{equation}
\hl{If all mutations of the antigen $a$ are equally likely and the affinity score $s$ was a nice linear function, then the solution to }(\ref{eqn:opt}) \hl{might be the same as } (\ref{eqn:plasmaOpt})\hl{. But we expect there to be substantive asymmetry in the probability of different antigen mutations, so we expect the solution to } (\ref{eqn:opt}) \hl{will be different than the solution to } (\ref{eqn:plasmaOpt})\hl{. This is the same principle as in the famous Wayne Gretzky quote about hockey, \emph{I skate to where the puck is going to be, not where it has been}.}

\hl{However, the situation is more complex, because in each germinal center} AM actually produces a \emph{set} of $N$ memory B cells.  We hypothesize that AM is evolved to try to produce a \emph{diverse} set of $N$ memory B cells that \hl{maximizes the expected affinity between the best-fit of the $N$ memory B cells and the random mutated antigen $\tilde{A}$}:
\begin{equation}
    \argmax_{ \{b_n \in \mathcal{B}, n=1, \ldots, N \} } \: \: E_{\tilde{A}} \bigg[  \max_{n=1,\ldots N} \left[  s(b_n, \tilde{A}) \right] \bigg].
\label{eqn:setoptmax}
\end{equation}

\begin{figure}
    \centering
    \tcbox[colback=white,top=5pt,left=5pt,right=5pt,bottom=5pt,colframe=mygray]{\includegraphics[width=.55\textwidth]{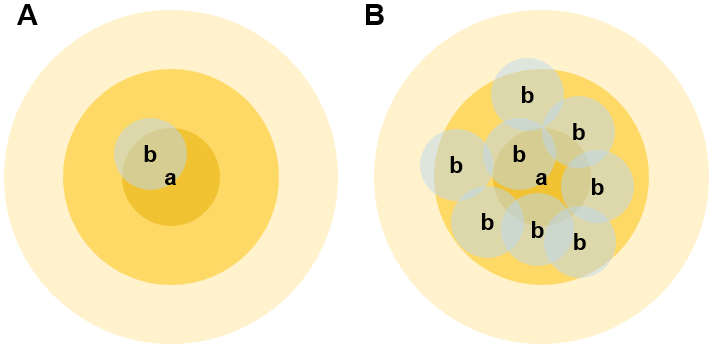}}
    \caption{\hl{Memory B cell coverage of the antigen mutation space. Given an antigen $a$, we expect to see a mutated version $\tilde{a}$ in the future. Suppose we cannot predict which mutations are more likely, so the probability distribution of the future mutated antigen $\tilde{a}$ is symmetric in the mutation space, shown here as yellow rings of decreasing probability from the original antigen $a$. (A) If you only get to choose one memory cell, and the antigen is equally likely to mutate randomly in any way, it is optimal to be a copy of a very good plasma cell, marked by $b$. The blue circle shows the likely mutations of $b$ after its $K$ mutations in a secondary germinal center. (B) If you get to keep a set of $N$ memory cells, spreading the memory B cells out will lead to a higher affinity to  more of the possible antigen mutations, rather than keeping $N$ copies of the plasma B cells.}}
    \label{fig:antigen_mut}
\end{figure}

Fig. \ref{fig:antigen_mut} illustrates the criterion in (\ref{eqn:setoptmax}), showing that diversity in the memory B cells helps cover the space of probable mutations of the original antigen $a$. 

The criterion (\ref{eqn:setoptmax}) assumes a fixed choice of $N$, but if you could also optimize (\ref{eqn:setoptmax}) for $N$, then you would always prefer a larger number $N$ of memory B cells. However, there is also downward pressure on $N$ due to the physical resources needed to store and maintain those cells, and time pressure before the antigen decays away.

\hl{We emphasize again that we are not hypothesizing that the memory B cells themselves try to optimize } (\ref{eqn:setoptmax})\hl{, rather that evolutionary pressures would have preferred memory B cell generation processes that optimized }(\ref{eqn:setoptmax})\hl{.}

\subsection{Optimizing For Warm Starting Training For A Mutated Antigen}
\hl{A second role of memory B cells is to warm start future AM processes for $\tilde{a}$. We argue that this role calls for a different criteria as to what makes for a good set of memory B cells. Specifically, analogous to } (\ref{eqn:minExpectedNumMutationsPlasmaCells})\hl{, we hypothesize that} the set of $N$ memory B cells $\{b_n \in \mathcal{B}\}$ should be chosen to minimize the number \hl{$K$ of mutations in the secondary response needed to produce a set of $N$ randomly mutated B cells $\{M^K(b_n)\}$ such that one of them is expected to become a secondary response plasma B cell, that is, that it meets the} affinity threshold $\tau$ with respect to the randomly mutated antigen $\tilde{A}$:
\begin{equation}
    \argmin_{K \in \mathbb{N}, \{b_n \in \mathcal{B}, n=1, \ldots, N \}}  \: \:   \left( K 
    \textrm{ subject to } \left( E_{\tilde{A}} \Bigg[  E_{M^K}\bigg[\max_{n=1, \ldots, N} s(M^K(b_n), \tilde{A})\bigg] \Bigg] \right) \geq \tau \right). 
    \label{eqn:setMinExpectedNumMutations}
\end{equation}

\hl{We do not mean to suggest that the memory B cells directly optimize }(\ref{eqn:setMinExpectedNumMutations})\hl{, but rather that evolutionary pressures might have preferred memory B cell selection processes that better optimize } (\ref{eqn:setMinExpectedNumMutations})\hl{.} 
  
Goal (\ref{eqn:setoptmax}) and goal (\ref{eqn:setMinExpectedNumMutations}) will probably have different optimal solutions depending on the probability of different mutations of the antigen and the B cells, though the same heuristic memory B cell selection process might do pretty well at both \hl{objectives.} It is not yet known how important memory B cells are to the secondary response plasma B cell training in germinal centers, yet some evidence shows that secondary response germinal centers are comprised of more
naive B cells than one might expect \cite{mesin2020restricted}. 

\subsection{Why Are Memory B Cells Not Copies Of Plasma B Cells?}
Both (\ref{eqn:setoptmax}) and (\ref{eqn:setMinExpectedNumMutations})  \emph{appear} to require knowledge of the probability distribution $P_{\tilde{A}}$ of different mutations the antigen may undergo, and the probability distribution $P_{M^K(b)}$ of a mutated B cell after $K$ mutations. However, for many symmetric choices of $P_{\tilde{A}}$ and $P_{M^K(b)}$, the exact distributions might not matter much: the immune system could cheaply achieve a good approximate solution to (\ref{eqn:setoptmax}) and (\ref{eqn:setMinExpectedNumMutations}) by simply making the memory B cells copies of the plasma B cells; there does not appear to be any biochemical restriction preventing exact replication. However, memory B cells do in fact appear to be selected for differently than the plasma B cells. We present two hypotheses as to why. 

\hl{Our first hypothesis was already introduced in Section } \ref{sec:diversitygood} \hl{and Fig. }\ref{fig:antigen_mut}\hl{: because the immune system gets to select}  a \emph{set} of memory B cells in (\ref{eqn:setoptmax}), it pays to have more diversity in the memory B cells than one gets by copying the plasma B cells. Plasma B cells tend to be less diverse because they are trained to optimize (\ref{eqn:plasmaOpt}), which even with multiple local minima \hl{in the shape of $s(b,a)$}, will limit their diversity. Memory B cells are more diverse than plasma B cells because they \hl{are selected} earlier in the maturation process. We believe this diversity is important to optimize affinity to the mutated antigen as per (\ref{eqn:setMinExpectedNumMutations}) because the true affinity function $s$ is a highly nonlinear function of the amino acid sequences of a B cell $b$ and antigen $a$ that arise from complex biochemical properties and physical lock-and-key structures \cite{CarneiroStewart:1994,Kilambi:2017,Ambrosetti:2020}.

Our second hypothesis is that the warm start \hl{objective} (\ref{eqn:setMinExpectedNumMutations}) for the secondary germinal centers is not well-optimized by a copy of the plasma B cell set because there is evidence that the probabilities $P_{M|b}$ of the random mutations of the B cells in SHM are \emph{asymmetric}: certain mutations of B cells are much more likely than others. That makes some B cells a more flexible starting point for warm-starting \hl{than the original plasma B cells, which may have trouble mutating to match the new antigen.}  Evidence for asymmetric $P_{M|b}$ is that many researchers have noted AID preferential targeting of specific motifs \cite{martin2015somatic, stavnezer2011complex, chen2017preferred, keim2013regulation}. As mutations would, by definition, change the specific sequence that AID targets, it is reasonable to infer that the first mutation of this location is easier than future ones. Once the sequence is changed, AID is less likely to target this location. Overall, the preferential targeting of AID would make it harder for this region to mutate further or reverse back to the original sequence. 

This asymmetry in the probability of moving around the space of all B cells via mutations during AM creates a disconnect between the plasma B cell \hl{objective} (\ref{eqn:plasmaOpt}) and the \hl{objective} of being a good warm start solution to future plasma B cell training as per (\ref{eqn:setMinExpectedNumMutations}). Specifically, a plasma B cell might have made many difficult-to-reverse mutations to optimize (\ref{eqn:plasmaOpt}) for the current antigen $a$. In contrast, the chosen memory B cells appear to be under-optimized for fitting the current antigen $a$, \hl{but we hypothesize they can more easily mutate in a secondary response AM to better fit} the random future antigen $\tilde{A}$. Overall, we note that how well the \hl{objectives} (\ref{eqn:plasmaOpt}), (\ref{eqn:setoptmax}) and  (\ref{eqn:setMinExpectedNumMutations}) align depends on the symmetry of $P_{\tilde{A}}$, $P_{M|b}$, and the nonlinearity of $s$. 

\section{Simulations} \label{sec:simulations}
We use the AM algorithm (Algorithm \ref{alg:shm}) to model how plasma B cells and memory B cells are trained, and show through two simulations that the simulated memory B cells are better than the simulated plasma B cells at optimizing our hypothesized \hl{objectives} (\ref{eqn:setoptmax}) and (\ref{eqn:setMinExpectedNumMutations}), thus providing evidence \hl{that these objectives are biologically reasonable. We first demonstrate the mechanics of affinity maturation in simulated primary immune responses, then compare different potential initial conditions for simulated secondary immune responses.} 

These simulations do not account for all of the real-world issues at play, such as that a viral mutation must not harm the virus's functionality, and the issues described in \ref{sec:knownIssues}. Despite these limitations, we argue these simulations capture many of the key issues needed to illustrate that our hypothesized \hl{objectives} are consistent with the difference in plasma and memory B cell training. Complete code for our simulations will be made available upon request.

\subsection{Simulation Set-up}
Our simulations follow  Algorithm \ref{alg:shm} for the AM process. We initialize a naive B cell repertoire (10,000 cells) with B cell receptors that are \hl{represented by} a random sequence of 10-50 amino acids, where each amino acid is drawn uniformly over the space of 61 non-stop codons (creating a non-uniform distribution over the amino acids). \hl{We simulate the antigens as sequences derived from known antigenic sequences of chicken ovalbumin, bovine milk, and wheat }\cite{honma1996allergenic,liu2018food}\hl{. Each antigenic sequence is 17 amino acids long for consistency.} We simulate the affinity metric $s$ between a B cell and the antigen using the standard \Verb vlocalalignv MATLAB function, which finds the optimal alignment between two sequences using the \Verb vBLOSUM50v matrix and returns a score reflecting how similar two sequences are in this alignment \cite{henikoff1992amino}. We use this score as a measure of affinity between the B cell receptor and the antigen for simplicity, but note that it is only a rough approximation of the more complex structural compatibility between a B cell receptor and an antigen.

Mutations of the B cell during AM are modeled in the codon space, where codons of the B cell receptor are replaced with one of the 61 codon possibilities. While SHM mutates B cells on a single nucleotide level, working in the codon space prevents the added complication of filtering out nonsensical B cell receptor sequences. In addition to the replacement of codons, codons in B cell receptor sequences can be inserted or deleted. \hl{Insertions and deletions are less likely to occur than swapping for another codon, based on rates of each type of mutation observed in humans} \cite{gibbs2003international,10002015global,zia2011ranking}\hl{.} The codons defining each B cell receptor are chosen to mutate at random, but we simulate codons that contain $C$ cytosines to be $C+1$ times more likely to be mutated than codons without cytosine, reflecting biological biases to nucleotide sequence motifs \cite{martin2015somatic, stavnezer2011complex, chen2017preferred, keim2013regulation}. \hl{We also impose a transition bias between codons, making some swaps more likely than others based on a mutability matrix derived from the} \Verb vBLOSUM50v \hl{matrix} \cite{henikoff1992amino,veerassamy2003transition}\hl{.}

For our initial simulation of a primary adaptive immune response to an antigen (primary response), we randomly select 50 naive B cells from the B cell repertoire from the top 1,000 of the 10,000 naive B cell repertoire in terms of affinity to said antigen. This reflects the recruitment of naive B cells with some partial affinity by T\textsubscript{FH} cells to germinal centers \cite{mesin2016germinal, tas2016visualizing}. These 50 `founder' B cells are then duplicated 20 times to form a germinal center population of 1,000 cells, reflecting the growth period of germinal center formation \cite{amitai2017population}. \hl{For each iteration of the simulation, 50 B cells are selected for proliferation and 50 B cells are selected for removal. Higher affinity B cells have a higher selection probability for proliferation, while lower affinity B cells have a higher selection probability for removal. The B cells selected for proliferation are duplicated and mutated, replacing all B cells selected during this iteration. This process imitates apoptosis of low affinity B cells from lack of T}\textsubscript{FH}\hl{ cell help and proliferation of B cells with high affinity after being selected by T}\textsubscript{FH}\hl{ cells.} These mutations have the possibility to increase, decrease, or have no effect on the affinity of the B cell receptors. \hl{We also impose constraints on which mutations can occur on a particular iteration, simulating a negative selection process due to damaging or potentially dangerous mutations. This entire process repeats over 100 iterations.}

As a basis for both Simulations 1 and 2, we extract 50 cells during the first half of our primary response simulation as the simulated memory B cells. The cells are randomly chosen from the top 25\textsuperscript{th} percentile of germinal center B cells. Another 50 cells are selected at the very end of the primary response to represent plasma B cells, where these cells exhibit the highest 50 affinity scores to the antigen. \hl{This reflects a slight deviation from Algorithm }\ref{alg:shm}\hl{, as we do not know the threshold $\tau$ \textit{a priori}. We establish $\tau$ for the secondary response based on the affinity of the plasma cells chosen at the end of the primary response.} As expected, the simulated memory B cells have overall lower affinity to the antigen compared to the plasma B cells, but higher than the initial set of naive B cells.

Our simulated mutations of the antigen for the secondary response are derived from a uniform random swap of any of the codons for any other (including possibly itself, i.e. a no-op), which creates a non-uniform distribution over the amino acids as some amino acids are coded for by multiple codons. \hl{Mutations of the antigen can also include insertions or deletions of codons, with equal probability to any codon.}

\subsection{Simulation 1: Affinity To A Mutated Antigen}

A secondary infection could involve an antigen that has been mutated or an antigen that is similar from a related pathogen. We simulate the changes in the antigen from the primary to secondary infection by causing adversarial mutations to our antigenic sequence. \hl{First, we generate 1,000 uniformly random mutations of antigen $a$'s sequence. The random mutations may be a swap of any amino acid to any other, an insertion of any amino acid, or a deletion of an amino acid.} Of those candidate mutations, we keep the one that has the lowest average affinity (worst case) to the set of $50$ plasma B cells from the primary response, to reflect that a potentially dangerous secondary infection would likely be from a more challenging mutation. We repeat this random process a total of $K$ times to produce an antigen with $K$ mutations.  \hl{We take that \emph{adversarial} antigen $\tilde{a}$ to be the worst case realization of $P_{\tilde{A}}$ from the candidate mutations.}

\begin{figure}[H]
    \centering
    \includegraphics[width=\textwidth]{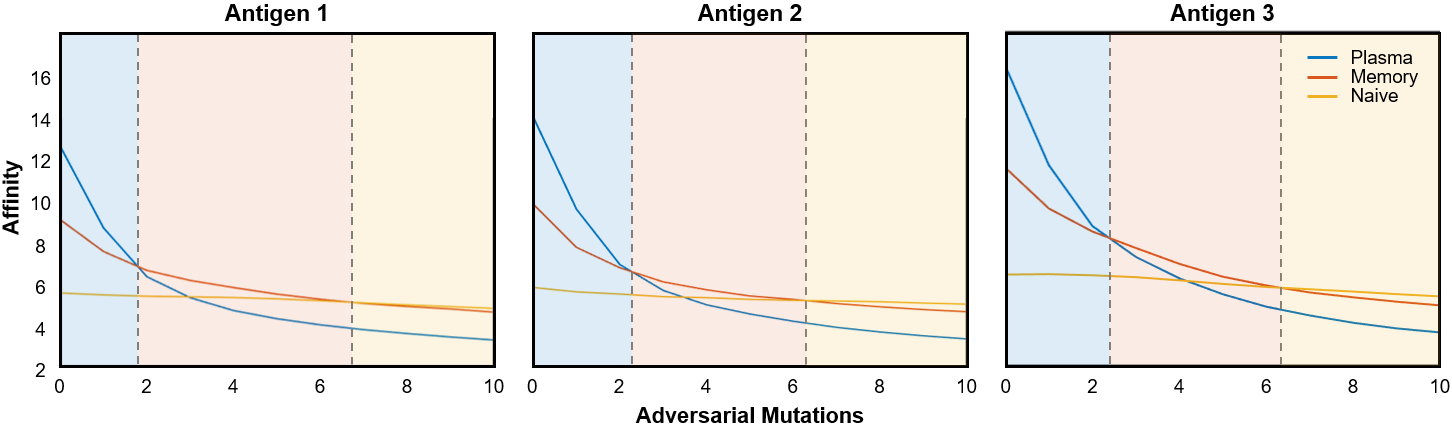}
    \caption{\hl{Potential initial conditions for a secondary adaptive immune response.} Simulation results averaged over 100 independent runs of the \hl{primary response for three antigens (from left to right: chicken ovalbumin, bovine milk, and wheat)}. Plots show average affinity of naive, plasma, and memory B cells to adversarially mutated antigens with $K = 1,\dots,10$ mutations. Ranges of mutations where plasma, memory, and naive B cells are optimal are shaded blue, orange, and yellow respectively.}
    \label{fig:hyp_1}  
\end{figure}

We then test the different B cell populations by the hypothesized goal of (\ref{eqn:setoptmax}). \hl{Fig. }\ref{fig:hyp_1}\hl{ shows the average affinity between increasingly mutated antigens (chicken ovalbumin, bovine milk, and wheat) with mutations $K = 1, \ldots,10$ for different cell populations at the end of the primary response, averaged over 100 independent runs } \cite{honma1996allergenic,liu2018food}. Fig. \ref{fig:hyp_1} shows that \hl{our simulated} plasma B cells are the best choice to maximize (\ref{eqn:setoptmax}) for a small number of adversarial mutations, \hl{our simulated} memory B cells are the best choice between $\sim$2-6 mutations, and naive B cells are best after many mutations (7+). Our simulations are too simplified for the specific transition points to be meaningful, but we argue they do provide strong evidence that the plasma B cells are likely not optimal for identifying substantially-mutated antigens. \hl{This may further suggest that there is some region in mutation space where memory B cells are more useful than plasma B cells or naive B cells as an initial condition for a secondary response.} 

We hypothesize that the fact that plasma B cells do not always have the highest affinity to mutated antigens is driven by the greater diversity of the memory B cells and naive B cells. While the mutations occur in DNA space, the relevant diversity is in the resulting nonlinear physical and biochemical space that defines the affinity to the antigen. \hl{Approximately measuring the B cell diversity in each population} using the pairwise BLOSUM similarities in each set shows substantial diversity differences, with the plasma B cells having average within-set BLOSUM similarity of $\sim$46, the memory B cells having much lower average within-set BLOSUM similarity of $\sim$27, and the naive B cells having even lower within-set BLOSUM similarity of $\sim$9 (from a representative primary response). 

\subsection{Simulation 2: Mutations Needed For Secondary Response Plasma B Cell Training}
In this simulation, \hl{we compare how well the three types of B cells perform as seeds for the secondary response germinal center.} We mimic a secondary response training of a new set of plasma B cell's optimized for high affinity to a mutated antigen $\tilde{a}$ (described in Simulation 1).  We initialize the secondary response plasma B cell optimization with one of three choices: \emph{(i)} the $N=50$ plasma B cells generated in the primary response for the original antigen $a$, \emph{(ii)} the $N=50$ memory B cells generated in the primary response for the original antigen $a$, or \emph{(iii)} $N=50$ naive B cells. We populate the germinal center in an identical way to the primary response, using the three sets of 50 B cells as our new founder cells. As in the \hl{primary response,} the naive B cells are random, but chosen to have some initial affinity to the now-mutated antigen to simulate recruitment to the germinal center. Each of these three secondary response germinal centers undergo AM in an identical fashion to the primary response. 

\begin{figure}[H]
    \centering
    \includegraphics[width=\textwidth]{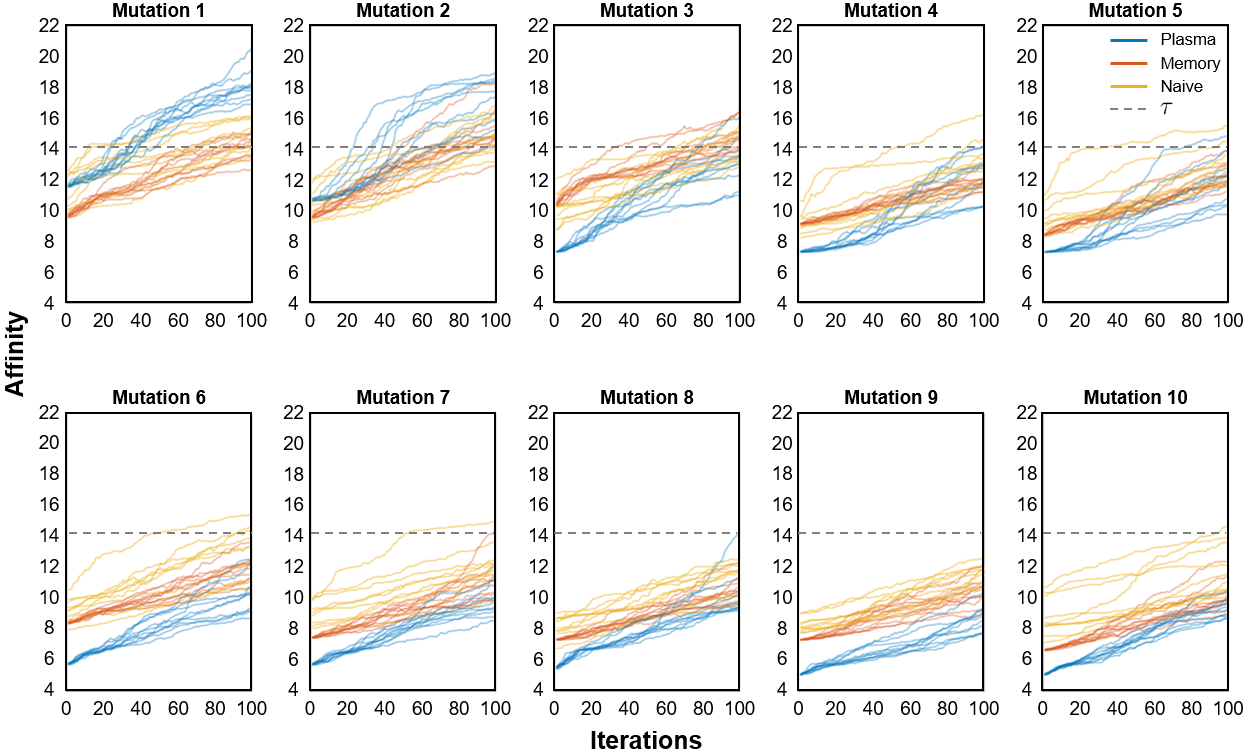}
    \caption{\hl{Example performance of different populations as initial conditions for secondary responses. We chose one representative run of the primary response simulation to the bovine milk antigen }\cite{liu2018food}. \hl{Each line represents one of the 10 secondary response simulations as a function of SHM iterations when initialized by a set of 50 plasma, memory, or naive B cells (as marked), for $K=1, \dots, 10$ sequential adversarial mutations. The y-axis marks the average affinity of the top 50 (out of 1,000) highest affinity B cells during the secondary response to the mutated antigen, representing the new potential plasma B cells from the secondary response.  Dashed horizontal line reflects the average affinity, $\tau$, of the plasma cells from the primary response. For this run of the simulation, the plasma B cells had difficulty achieving high affinity to the mutated antigens after three adversarial mutations. The memory B cells were the most useful seeds for the three-mutation case. Once the antigen was mutated four times, only the naive B cells were sometimes able to achieve the affinity threshold before 100 iterations of SHM. Surprisingly, the average affinities plotted grew fairly linearly and at roughly the same rate for most of the lines, suggesting fairly constant progress, and that none of the populations became stuck, but rather just started from a worse initial affinity.}}
    \label{fig:hyp_2}
\end{figure}

\begin{figure}[H]
    \centering
    \includegraphics[width=.75\textwidth]{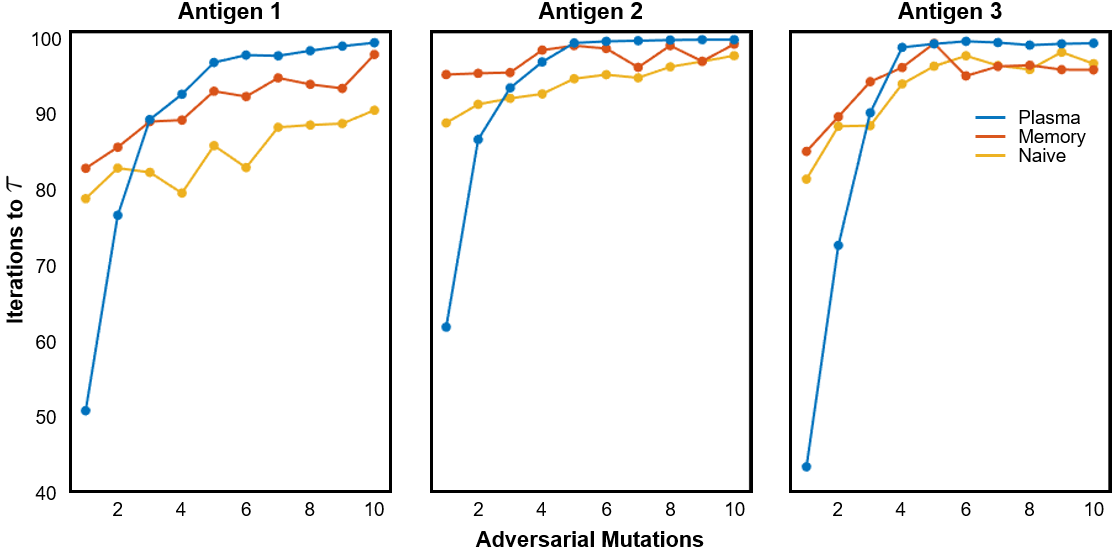}
    \caption{\hl{Convergence rate for potential initial conditions across all secondary responses.} Time of convergence to $\tau$ for plasma, memory, and naive B cells during simulated secondary germinal centers' warm starts for $K = 1,\dots,10$  mutations \hl{averaged over 50 independent runs of the entire simulation, where each of the primary response simulations had 10 corresponding secondary response simulations.} Lower values indicate a faster time to convergence, showing that the plasma B cells are the best warm starts for \hl{one or two} of the simulated adversarial mutations (dependent on the antigen). \hl{Naive B cells are the best warm starts for almost all cases with three or more simulated adversarial mutations. Antigens 1, 2, and 3 correspond to chicken ovalbumin, bovine milk, and wheat, respectively }\cite{honma1996allergenic,liu2018food}.}
    \label{fig:hyp_2_tau}
\end{figure}

\hl{Fig. }\ref{fig:hyp_2}\hl{ are representative examples of multiple secondary response simulations, given the same primary response and the same $K=1, \dots, 10$ mutations on the bovine milk antigen }\cite{liu2018food}\hl{. Specifically, it shows the average affinity of the 0.05\% highest affinity cells over 100 iterations in secondary responses. Fig. }\ref{fig:hyp_2}\hl{ highlights that for just one or two antigen mutations, plasma B cells tend have the highest initial affinity to the mutated antigen. For three or more adversarial mutations, naive and memory B cells are better initial conditions for the secondary response than the primary response plasma B cells, and reach an affinity of $\tau$ in fewer iterations (Figs. }\ref{fig:hyp_2}\hl{ and }\ref{fig:hyp_2_tau}\hl{). Fig. } \ref{fig:hyp_2}\hl{ also conveys how the recruited secondary response naive B cells tend to have higher initial affinity to the mutated antigen than either the primary response plasma B cell or memory B cells once the antigen has been sufficiently mutated. This was expected, but we were surprised at how few mutations it took for the naive B cells to have the highest affinity at the first iteration of the secondary response. That is, for relatively few codon mutations in the new antigen, we find empirical evidence that seeding a secondary response with primarily B cells from the naive repertoire is an advantageous strategy. We were not able to simulate a region in mutation space where the memory B cells from the primary response consistently had the highest affinity at the first iteration of the secondary response, likely due to considerable variation between simulated secondary responses.}

\hl{Similarly, Fig. }\ref{fig:hyp_2_tau}\hl{ shows the average number of iterations for plasma, memory, and naive B cells to reach $\tau$ over all secondary response simulations for all three antigens. The iteration number, where the threshold $\tau$ is reached, is averaged over all secondary response simulations for 50 independent simulations of the primary response for each antigen. Again, our simulations did not show a region of mutation space where memory B cells were consistently the best warm start conditions. We found this to persist across a number of refinements of our simulation, suggesting this finding might not be an artifact of too coarse a simulation.} We hypothesize that this is due to the selection of naive B cells for the secondary germinal center being biased to have some initial affinity to the mutated antigen (as described in the primary response simulation). \hl{That is, naive B cells in Fig. }\ref{fig:hyp_2}\hl{ are selected for initial affinity to $\tilde{a}$, while memory and plasma B cells derived from naive B cells selected for having some initial affinity to the original antigen $a$ (Fig. }\ref{fig:hyp_1})\hl{. We initially suspected that our affinity bias for the selection of naive cells was too strong, but both naive and memory B cells are recruited simultaneously in real secondary germinal centers in order to have the best chance of creating new plasma B cells.} In fact, recent experimental evidence suggests that secondary response germinal centers are comprised of more naive B cells than previously thought \cite{mesin2020restricted}. It is also possible that there is an antigen mutation regime for which the memory B cells are indeed more effective than naive B cells for warm starting, but that our simulations were not realistic enough to capture it.  

\hl{We note that the secondary response simulations often take more than 100 iterations to reach the affinity threshold $\tau$ from their corresponding primary simulation. When calculating the average number of iterations to reach $\tau$ in Fig. } \ref{fig:hyp_2_tau}\hl{, we set these cases to the maximum iteration number of 100. We hypothesize that this phenomenon occurs because the adversarial mutations of $\tilde{a}$ may create a more difficult problem for the simulated germinal center to solve.}

\section{Conclusions and Open Questions} \label{sec:conclusions}
We hypothesized that the dual role of memory B cells can be captured by two objectives (\ref{eqn:setoptmax}) and (\ref{eqn:setMinExpectedNumMutations}), and that these \hl{objectives} would not be as well-optimized by copying plasma B cells that are trained for (\ref{eqn:plasmaOpt}), due to their over-fitting the original antigen.  Our simulations, while limited, provide strong evidence that plasma B cells would not optimize (\ref{eqn:setoptmax}) or (\ref{eqn:setMinExpectedNumMutations}) once the antigen underwent sufficient adversarial mutations. We believe this suboptimality of plasma B cells against mutated antigens provides a role for the different selection mechanism used for memory B cells.

Our simulations show a limited range of antigen mutations over which our simulated memory B cells may be optimal; for substantial mutations, we show random naive B cells can work even better. These findings are consistent with our knowledge of the adaptive immune system. Plasma B cells are a one-time solution and are highly overfit to the current antigen of interest. Memory B cells provide a more approximate solution to the current antigen, which is kept within the body to recognize future antigens with similar characteristics. If a future antigen is so different from what has been previously encountered by the immune system that no memory B cells are able to identify it, a new solution is formed from scratch using naive B cells.

Memory B cells play two roles, both differentiating into plasma B cells and re-initiating germinal centers, but these roles may be played by distinct subpopulations \cite{Dogan:2009,Weisel:2017}. These distinct subpopulations of memory B cells might have resulted from distinct AM processes, or changes in the AM process over the AM time span that we have not explicitly modeled in our Algorithm \ref{alg:shm} \cite{Weisel:2016}. Thus our two memory B cell objectives in (\ref{eqn:setoptmax}) and (\ref{eqn:setMinExpectedNumMutations}) may apply to independent memory B cell populations. Here we investigated a simplified model, where memory B cells were considered a unified group. However, we were not able to show via simulations a regime in which the memory B cells were clearly better than secondary naive B cells for re-initiating the secondary germinal response. Our results suggest this re-initialization task might be a weaker or rarer role of the memory B cells. \hl{These results align with} recent experimental evidence that similarly noted memory B cells were less prevalent in secondary germinal centers than previously assumed \cite{mesin2020restricted}. However, even if memory B cells are not always needed for the secondary response, it might be that in some cases they are very important for warm starting, which might still exert evolutionary pressure on their selection process.

The evolutionary pressures on memory B cell selection in nature are not known, but may be elucidated through the integration of computational simulations and biological experiments. Actively monitoring the affinity of B cells during affinity maturation, as well as detecting when and why GC B cells become memory B cells, may assist with the development of more accurate and complex models in the future.  

\section{Acknowledgements}
We thank Roger Brockett, Alnawaz Rehemtulla, Serena Wang, Santosh Srivastava, Charles Ryan, Sijia Liu, Ren Wang, Tianqi Chen, and Christopher York for feedback on the manuscript and helpful discussions. This work is supported by the Guaranteeing AI Robustness against Deception (GARD) program from DARPA/I2O.

\clearpage
\bibliography{references}

\end{document}